\documentclass[aps,prl,twocolumn,groupedaddress]{revtex4}
\def\slash{{\kern 2pt\raise 1.5pt \hbox{$\backslash$} \kern -9pt}}

\def\t{\tau}

\def\beq{\begin{equation}}
\def\eeq{\end{equation}}
\def\bea{\begin{eqnarray}}
\def\eea{\end{eqnarray}}

\begin{document}
\eprint{DIAS-STP-10-01}
\title{Chiral fermions and torsion in the early Universe}
\author{Brian P. Dolan}
\affiliation{Dept.~of~Mathematical~Physics,~NUI~Maynooth,~Ireland\\
{\rm and}\\
School of Theoretical Physics, Dublin Institute for Advanced Studies,
10~Burlington Rd., Dublin 8, Ireland}
\email{bdolan@thphys.nuim.ie}

\begin{abstract}

Torsion arising from fermionic matter in the Einstein-Cartan formulation 
of general relativity is considered in the context of Robertson-Walker
geometries and the early Universe.  An ambiguity in the way
torsion arising from hot fermionic matter in chiral models
should be implemented is 
highlighted and discussed. In one interpretation, chemical 
potentials in chiral models can contribute to the Friedmann 
equation and give a negative contribution to the
energy density.


\end{abstract}

\maketitle

It is often the case that quantum matter acts as a source for a classical
field in situations where quantum aspects of the field
itself can be ignored.  
This approximation has proven extremely useful for Einstein's equations where
\beq \label{Einstein}
G_{ab}=8\pi G <T_{ab}>,
\eeq
works well when the matter source is degenerate fermionic matter, 
where $<>$ is a quantum expectation value, and for thermal
radiation, where $<>$ is a thermal average of photons.
There are difficulties with this approach however, 
not least that the singularities inherent in fully fledged
quantum field theory for the sources render
(\ref{Einstein}) ambiguous and some criterion for
cutting off the integrals must be introduced.
For example it is well known that
a na{\rm \" i}ve calculation of  the vacuum energy density of the standard
model of particle physics
leads to far too high a value of the cosmological constant to 
be compatible with observations \cite{Weinberg1}. Nevertheless (\ref{Einstein}) seems to
work well in the early
Universe when the dynamics is dominated by radiation, 
as long as temperatures are well below the Planck temperature.  
In the radiation dominated Universe Einstein's
equations boil down to the Friedmann-Robertson-Walker (FRW)
equation, ignoring spatial curvature this is 
\beq
\frac{\dot a^2}{a^2}=\frac{8\pi G} 3  <T_{00}> =
 N_{eff} \frac {4\pi^3} {45}\frac{T^4}{m_{Pl}^2} 
\label{StandardFRW}
\eeq
where $a(t)$ is the cosmological scale factor and
$N_{eff}$ is the effective number of degrees of freedom in the
relativistic gas, 
\beq N_{eff}:=N_B +\frac 7 8 (N_+ + N_-)
\eeq
with $N_B$ the number of bosonic degrees of freedom (2 for photons),
and $N_+$ and $N_-$
the number of positive chirality and negative chirality 
fermionic degrees of freedom respectively (for a standard model neutrino
$N_+=2$, $N_-=0$; for a Dirac fermion $N_+=N_-=2$), \cite{Weinberg2}.
The Planck mass, $m_{Pl}^2=G^{-1}$ (we use units with $\hbar=c=1$), 
appears in (\ref{StandardFRW})
not because we are considering a theory of quantum gravity but because
of the quantum nature of the source for classical gravity.

In the Einstein-Cartan formulation of general relativity
fermionic matter is expected to induce torsion (recent bounds on
the magnitude of torsion have been derived from
tests of violation of Lorentz invariance \cite{KRT} and from
cosmic microwave polarization \cite {DasMohantyPrasanna}). 
When the connection is varied in the Einstein-Cartan action
the torsion two-forms $\t^a=\frac 1 2 \t^a{}_{bc} e^b\wedge e^c$
are determined by a spinor field $\Psi$ via the algebraic equation 
\beq \label{Torsion}
\t^a=2\pi  G\epsilon_{abcd}(\overline\Psi \gamma^5\gamma^d \Psi)e^b\wedge e^c,
\eeq
$a,b,c,\ldots$ are orthonormal indices (for a review of torsion in
Einstein-Cartan formulation in general see \cite{Hehl} and \cite{Shapiro}).

In the spirit of (\ref{Einstein}) the equation of motion
(\ref{Torsion}) would be interpreted as
\beq \label{TorsionThermalAverage}
\t_{a,bc}= 4\pi G\epsilon_{abcd}<\overline\Psi \gamma^5\gamma^d \Psi>
=-4\pi G\epsilon_{abcd}<j^d_5>,
\eeq
where $j^a_5=\overline\Psi \gamma^a\gamma^5 \Psi$ is the axial current.
As is well known fermions generate
torsion in the anti-symmetric class of tensors,
according to the classification of \cite{CLS}.

We shall examine the effect of torsion arising from
relativistic fermions in the early
Universe, assuming isotropy and spatial homogeneity of both
the geometry and the matter.  
It will be assumed that the metric of Robertson-Walker type 
and that the energy-momentum is of the form 
\beq T_{ab}=\left(\begin{array}{cc} \rho & 0 \\ 0 & P\, \delta_{ij} \end{array}
\right) \eeq
where the density $\rho$ and pressure $P$ are homogeneous and depend only
on time and $i,j=1,2,3$ are space-like indices.
  
The Riemann tensor involves
the square of the connection and 
the net effect of including the torsion (\ref{Torsion}) 
into the gravitational connection is
that Einstein's equations are modified to
\bea \label{EinsteinTime}
3\left(\frac{\dot a^2}{a^2} -\frac {\t^2} 4\right)  &=& 8\pi G  \rho \\
\label{EinsteinSpace}
 -\frac {2\ddot a} a -\frac{\dot a^2}{a^2} +\frac {\t^2} 4 
&=& 8\pi G P,
\eea
where $a(t)$ is the Robertson-Walker scale factor and
\beq \label{TorsionSquared}
\t^2=-\t^a\t_a=
-16\pi^2 G^2(\overline\Psi \gamma^5\gamma^a \Psi)(\overline\Psi \gamma^5\gamma_a \Psi)
\eeq
(the metric signature is $(-,+,+,+)$,
the Clifford algebra convention is $\{\gamma^a,\gamma^b\}=-2\eta^{ab}$,
with $\gamma^0$ hermitian, and spatial curavature is taken to be zero).
Equations (\ref{EinsteinTime}) and (\ref{EinsteinSpace})
are not independent and are related by the
(first) Bianchi identity which is analysed below.
Eliminating $\tau$ gives
the usual relation between the acceleration and the density and pressure,
\beq \label{Acceleration}
\frac{\ddot a} a = -\frac {4\pi G} 3 (\rho + 3P).
\eeq

Since fermions constitute quantum matter, it seems natural to
interpret (\ref{Torsion}) in the early Universe as meaning 
a thermal average (\ref{TorsionThermalAverage}).
However there is an ambiguity as to whether (\ref{TorsionSquared}) 
should be interpreted using
\beq \label{SpacelikePsiFour}
<(\overline\Psi \gamma^5\gamma^a \Psi)(\overline\Psi \gamma^5\gamma_a \Psi)>
=<j_5^a j_{5,a}>
\eeq
or
\beq \label{TimelikePsiFour}
<\overline\Psi \gamma^5\gamma^a \Psi><\overline\Psi \gamma^5\gamma_a \Psi>
=<j^a_5>< j_{5,a}>\eeq

These are different in general.  The former can be Fierz
re-arranged to give
\bea \label{Fierztau}
<(\overline\Psi \gamma^5\gamma^a \Psi)(\overline\Psi \gamma^5\gamma_a \Psi)>
&=&\hskip 80pt \  \nonumber \\ \label{Fierz}
4<(\Psi_+^\dagger&&\kern -20pt\Psi_-) (\Psi_-^\dagger\Psi_+)>
\eea
where $\Psi_+$ and $\Psi_-$ are the positive and negative chirality
components of $\Psi$.
In a first quantised theory this is positive definite for Dirac spinors
and vanishes for Weyl spinors \cite{Hehl},
hence $\t^2 < 0$ in (\ref{TorsionSquared}), making the torsion space-like
(in the classification of \cite{CLS} this is denoted {\it As}).
The cosmological consequences of this formulation in inflationary models
are explored in \cite{Watanabe}. 
The same philosophy, applied to spin densities  rather than the pseudo-vector
$\overline\Psi \gamma^5\gamma^a \Psi$, is followed in \cite{NurgalievPonomariev}
and \cite{Gasperini}. 

We reach a radically different conclusion if we use (\ref{TimelikePsiFour}),
which follows from taking the thermal average of (\ref{Torsion}) before calculating
the Riemann tensor.
Applying the usual Robertson-Walker 
assumptions of spatial homogeneity and isotropy to the connection, and hence the
torsion, we would conclude that, in the cosmic frame,
\beq
<j^i_5>=0,
\eeq
while
\beq \label{SMtau}
<j^0_5>=(n_+-\overline n_+) -(n_- - \overline n_-)
\eeq
where $n_+$, $\overline n_+$, $n_-$ and $\overline n_-$ are thermal
averages of the number densities of positive chirality particles, positive chirality 
anti-particles, negative chirality particles and negative chirality 
anti-particles respectively.
In contrast the vector current $j^a=\overline\Psi \gamma^a \Psi)$
has
\beq
<j^0>=(n_+-\overline n_+) +(n_- - \overline n_-)=n-\overline n,
\eeq
where $n=n_++n_-$ is the number density of all particles and $\overline n=
\overline n_+ + \overline n_-$ the number density of anti-particles.
The integral of $j^0$ over a co-moving 3-dimensional volume $V$ gives the
total number of fermions in that volume, counting anti-particles as
negative,
\beq\int_V j^0 a^3(t)d^3x= {\cal N} - \overline {\cal N},\eeq
where ${\cal N}=\int_V n a^3(t)d^3x$, etc.
The corresponding number for the axial current,
\beq\int_V j^0_5 a^3(t)d^3x = {\cal N}_+ - {\cal N}_- -
(\overline {\cal N}_+-\overline {\cal N}-):={\cal N}_A 
\eeq
we shall call the axial particle number in the volume $V$.

In the abscence of chemical potentials, for a thermal distribution
with the temperature much greater than any particle masses, 
\beq n_\pm=\overline n_\pm=\frac 3 4 \frac {\zeta(3)} {\pi^2} T^3\eeq
so ${\cal N}=\overline {\cal N}$ and the torsion vanishes.
A non-zero chemical potential is necessary for any asymmetry between particle
and anti-particle numbers.  We see here that a chemical potential
can also generate torsion, since then 
\beq n_\pm-\overline n_\pm=\frac 1 6 \mu_\pm T^2+o\left(\frac {\mu_\pm}{T}\right)^2T^3\eeq
for $\mu_\pm << T$, 
where $\mu_+$ and $\mu_-$ are chemical potentials for positive 
and negative chirality particles, see {\it e.g.}~\cite{Weinberg2}.
In a chiral theory, such as the Standard Model, 
$\mu_+$ and $\mu_-$ can be different in general and
\beq < j^0_5 > = \frac 1 6 (\mu_+ - \mu_-)T^2,\eeq
where we have ignored terms $o\left(\frac {\mu_\pm}{T}\right)^2$.
In a Robertson-Walker Universe undergoing adiabatic expansion, $a\propto 1/T$,
the thermal average of the
axial current \hbox{$\nabla_a <j^a_5>=0$}
is conserved, and hence ${\cal N}_A$ is constant, 
if and only if $(\mu_+ - \mu_-) \propto T$ is linear in $T$.
The total fermion number, ${\cal N}-\overline {\cal N}$, 
is constant if and only if
$(\mu_+ + \mu_-) \propto T$ (particle masses are being ignored here).

To summarise, in general equations 
(\ref{TorsionThermalAverage}) and (\ref{TimelikePsiFour}) give
\beq
\t^2=\frac{4\pi^2 G^2}{9} (\mu_+-\mu_-)^2 T^4
\eeq
which is positive in any chiral model for matter with $\mu_+\ne \mu_-$.
For a model with $N_f$ different types of fermion each fermions
species can have different chemical potentials $\mu^{(k)}_\pm$, 
where we label the species
with an integer $k$, and then $\mu_\pm$ is always understood below
to mean $\mu_\pm=\sum_{k=1}^{N_f} \mu^{(k)}_\pm$.
Since positive and negative chirality particles in a chiral model can
have different weak charges one expects that $\mu^{(k)}_+ \ne \mu^{(k)}_-$
in general,
and the torsion can be non-zero, at least for the era before the electro-weak
phase transition, \cite{HT}.

Define
\beq \label{muT2}
\t=\frac {2\pi G}{3} (\mu_+-\mu_-) T^2,
\eeq
in terms of which the non-vanishing components of the torsion are
\beq 
\t_{i,jk}=\epsilon_{ijk}\t,\eeq
(in the classification of \cite{CLS} this is time-like, {\it At}).
Rotational invariance of the thermal average is
not incompatible with the conclusion 
of \cite{Kuchowicz}, where classical solutions of the Weyl equation
were analyzed in spherical symmetric space-times with torsion --- 
thermal averages do not necessarily have the same symmetries as
solutions of the equations of motion.
The general form of the torsion compatible with Robertson-Walker 
symmetries was given in \cite{Tsamparlis}.
The fact that chiral fermions can have interesting consequences
when torsion is taken into consideration was noted
in the context of anomalies for lepton currents in the
Standard Model of particle physics in \cite{DobadoMaroto}.

Both (\ref{SpacelikePsiFour}) and (\ref{TimelikePsiFour}) have interesting,
though very different, cosmological consequences.
The form (\ref{TimelikePsiFour}), being the square of
a vector, has a dual description as the square of a 3-form and as such
is in the class of models described in \cite{3Form}.  Indeed a term of this
form is present in the Landau-Ginsparg models discussed in \cite{3Form}, though
the stabilising quartic term is absent and there is
no kinetic term here.  
A kinetic term would require time derivatives
of the torsion and so would go beyond Einstein-Cartan theory --- such   
terms would be expected to appear in an effective action description 
of gravity involving higher derivatives and powers of the Riemann tensor.

So which should one use (\ref{SpacelikePsiFour}) or (\ref{TimelikePsiFour})?
Weinberg \cite{Weinberg3} takes the point of view that there is
nothing special about torsion: it is just another tensor and
one can always move it to the right hand side of Einstein's equations
and consider it to be part of the matter rather than part of the geometry.
We see here that, in the context of (\ref{Einstein}), there is an
ambiguity.  If the torsion terms
are absorbed into the energy momentum tensor before expectation
values are taken then it would seem that
(\ref{SpacelikePsiFour}) is appropriate.  In the Einstein-Cartan formulation
however the torsion is determined by the equation of motion (\ref{Torsion}),
in which the square of the torsion does not appear.
If the gravitational field itself is not quantised, it is hard to see
any interpretation of the equation of motion
(\ref{Torsion}) other than (\ref{TorsionThermalAverage}).
When the Riemann tensor is calculated it is then
(\ref{TimelikePsiFour}) that arises and not (\ref{SpacelikePsiFour}).
Much of the literature has focused on (\ref{SpacelikePsiFour}),
in a cosmological context for example
(\ref{SpacelikePsiFour}) was used in \cite{Watanabe}.
In this paper the consequences of (\ref{TorsionThermalAverage}) and
(\ref{TimelikePsiFour}) will be explored and developed.

We shall see that, in the context of the early Universe, 
the torsion can give a negative contribution to the energy density.
The mechanism here is different to
torsion induced avoidance of the initial singularity due to spin 
fluids considered
previously, \cite{Kopczynski,Trautman,Gasperini2,P-PF,AndradeRamos},
in which the spin density necessarily breaks either rotational or
translation invariance.

When there is torsion the Bianchi identity
does not require that $G_{ab}$ be co-variantly constant, in general
one has
\beq
\nabla_b G^{ba}=-\t^c{}_{bc}G^{ba}+\frac 1 2 \widetilde R^{adbc}\t_{d,bc},\eeq
where $\widetilde R^{adbc}:=\frac 1 4  \epsilon^{ada'd'}R_{a'd'b'c'}\epsilon^{bcb'c'}$.
In the case of Friedmann-Robertson-Walker (FRW)
Universes under study here only the second term on the right hand side contributes giving
\beq \label{DG}
\nabla_b G^{b0}=
-\frac 3 2 \frac \t a \frac d {dt} (\t a),\qquad
\nabla_b G^{bi}=0.
\eeq 
One strategy is to demand $\nabla_b G^{ba}= 0$ and use this to determine the
torsion, implying that $\t \propto 1/a$ \cite{Conserved-T-torsion}, but this is too restrictive for our purposes.  Instead we take thermodynamic averages
as above and use (\ref{muT2}) for the form of the torsion.
 
Assuming adiabatic expansion
$T\propto 1/a$, \cite{Weinberg2},  
$a$ can be eliminated from Einstein's equations 
in favour of $T$ to give
\bea \label{TdotEquation}
\frac{\dot T^2}{T^2}   &=& \frac{8\pi G} {3}  \rho
+ \frac { \pi^2 G^2 } {9} (\mu_+-\mu_-)^2 T^4, \\
\label{TddotEquation}
\frac {\ddot T} T
&=&  \frac{4\pi G}{3}  (5\rho +3 P)  + \frac {2 \pi^2 G^2} {9} (\mu_+ - \mu_-)^2 T^4.
\eea
These two equations are not independent, the Bianchi identity (\ref{DG})
gives an equation relating $\rho$, $P$, $T$ and $\mu_\pm$.
Expressing time derivatives as temperature derivatives, 
$\frac {d} {d t}= \dot T \frac {d} {d T}$, differentiating
(\ref{TdotEquation}) and using (\ref{TddotEquation}) to eliminate $\ddot T$
gives
\beq\label{BianchiEoS}
T^3\frac {d} {d T} \left\{(\mu_+-\mu_-)^2 T^2\right\} =\frac{24}{\pi G}
\left(3h - T\frac {d\rho} {d T}\right),
\eeq
where $h=\rho + P$ is the enthalpy density.
For standard equations of state (relativistic gas, dust,
cosmological constant) the right hand side of (\ref{BianchiEoS})
vanishes, so the torsion described here necessarily
requires a modification of the equation of state.
Let $\rho=\rho_0+\Delta\rho$, $P=P_0+\Delta P$ and $h=h_0+\Delta h$,
where $\rho_0$ etc., satisfy a standard equation of state. Then 
\beq\label{BianchiEoSDel}
T^3\frac {d} {d T} \left\{(\mu_+-\mu_-)^2 T^2\right\} =\frac{24}{\pi G}
\left(3\Delta h - T\frac {d \Delta \rho} {d T}\right).
\eeq

For a relativistic gas with the full energy-momentum tensor traceless 
$\Delta P=\frac{\Delta\rho} {3}$ and
(\ref{BianchiEoSDel}) reduces to
\beq
\frac {d} {d T} \left\{(\mu_+-\mu_-)^2 T^2\right\} =-\frac{24}{\pi G}
T^2\frac {d} {d T}\left(\frac{\Delta\rho}{T^4}\right).
\eeq

Two special cases are:

\begin{itemize}

\item If axial particle number is conserved, $(\mu_+ - \mu_-)=b T$ with
$b$ a constant,
so $\Delta\rho$ must have a $T^6$ component,
\beq \label{TorsionT3}
\rho=\frac{\pi^2}{30}N_{eff}T^4 -\frac{\pi G b^2}{12} T^6,
\eeq
where the $T^4$ term is assumed to have the usual form for massless
particles with $N_{eff}$ degrees of freedom.
In particular the torsion necessarily gives a negative contribution to
the energy density in this case, though in realistic models this
is a small effect.  We must take $b<<1$ to be consistent with our
assumption that $(\mu_+-\mu_-)<<T$.   This is compatible with
the observed value of the current ratio of Baryon to photon densities
\beq
\frac{n- \overline n}{n_\gamma} =  \frac{\pi^2}{12\zeta(3)}
\left(\frac{\mu_+ + \mu_-}{T}\right)\approx 10^{-9},
\eeq
with $n_\gamma=\frac {2\zeta(3)}{\pi ^2} T^3$.
Unless $\mu_+$ and $\mu_-$ have opposite sign 
and the smallness of this number
is due to a delicate cancellation between two large numbers, 
this requires $b$ to be of the order of, or less than, $10^{-9}$ 
at the present time.
Then the ratio of the two terms in
(\ref{TorsionT3}) is of order $\frac{b^2}{N_{eff}} \left(\frac {T }{m_{Pl}}\right)^2$ 
which is very small
for $T<<m_{Pl}$ (obviously the analysis here is only valid for $T<<m_{Pl}$
where quantum gravity effects are assumed to be very small).\newline
This behaviour $\t\sim 1/a^3$ for torsion 
arising from spin in the early Universe has been studied before,
\cite{Gasperini2,AndradeRamos}.

\item For any period during which $\mu_+ - \mu_-$ is constant, independent of temperature, equation (\ref{BianchiEoSDel}) gives a logarithmic correction to the
the energy density,
\beq\rho 
\approx \frac {\pi^2}{30} N_{eff} T^4 - \frac{\pi G}{12}(\mu_+ - \mu_-)^2T^4\ln \left(\frac {T}{T_0}\right),\eeq
where $T_0$ is an arbitrary constant that can be absorbed into the definition
of $N_{eff}$.  Axial particle number is not conserved in this case
and one expects Baryon/Lepton number violating processes
unless $\mu_+=-\mu_-$.

\end{itemize}

Of course the generation of 
Baryon or Lepton number requires $P$ and $CP$ violation as well 
as a period out of thermal equilibrium, according to the
Sakharov conditions, \cite{Sakharov}.  The details would depend
on the particular chiral model generating the chemical potentials 
and taking the Universe through a period during which it is out
of thermal equilibrium.

At first sight if might seem disconcerting that energy-momentum does
not appear to be conserved in this formalism --- because of (\ref{DG})
and the Einstein equations $T_{ab}$ 
cannot be co-variantly constant unless  $a\t$ is constant.  
However an ``improved''
energy-momentum tensor, which is conserved, can be defined.
We make the co-variant decomposition of the Einstein tensor
\beq
G_{ab}={\stackrel{0}{G}}_{ab} +\Delta G_{ab}
\eeq
where ${\stackrel{0}{G}}_{ab}$ is the Einstein tensor
constructed from the torsion-free connection.
We similarly decompose the connection one-forms as
\beq \label{ConnectionDecomposition}
\omega^a{}_b=\stackrel{0}{\omega}{}^a{}_b+\Delta\omega^a{}_b\eeq
with \hbox{$\stackrel{0}{\omega}\hskip -2pt {}^a{}_b$} the torsion-free connection. 
Expanding $\Delta\omega^a{}_b= \Delta\omega^a{}_{b,c}\,e^c$ 
the components $\Delta\omega^a{}_{b,c}$, being the difference of two connections,
constitute a tensor field
so (\ref{ConnectionDecomposition}) is again a co-variant decomposition.
${\stackrel{0}{G}}_{ab}$ is the zero torsion Einstein tensor 
for which the first Bianchi identity implies
\beq
\stackrel{0}{\nabla}_b \stackrel{0}{G}{}^{ba}=0
\eeq
where 
$\stackrel{0}{\nabla}_b$ is the co-variant derivative using 
\hbox{$\stackrel{0}{\omega} \kern -2pt {}^a{}_b$}.
From this follows
\beq
\nabla_b G^{ba}=\stackrel{0}{\nabla}_b\bigl(\Delta G^{ba}\bigr)
+ \Delta\omega^b{}_{c,b} G^{ca} + \Delta\omega^a{}_{c,b} G^{bc}.
\eeq
We also have, by definition,
\beq
\nabla_b T^{ba}=\stackrel{0}{\nabla}_b T^{ba}
+ \Delta\omega^b{}_{c,b} T^{ca} +
\Delta\omega^a{}_{c,b} T^{bc}
\eeq
for $T_{ab}$.
Einstein equations, $G^{ab}=8\pi G T^{ab}$, now imply
\beq
\stackrel{0}{\nabla}_b\bigl(\Delta G^{ba}\bigr)=8\pi G \stackrel{0}{\nabla}_b T^{ba}.
\eeq
An ``improved'' energy-momentum tensor can be defined
\beq {\cal T}^{ab}:= T^{ab}- \frac 1 {8\pi G}\Delta G^{ab}
\eeq
which is conserved using the torsion free connection,
\beq
\stackrel{0}{\nabla}_b {\cal T}^{ba}=0.
\eeq

For example if $\tau=\bigl(\frac {2\pi G b}{3}\bigr) T^3$ in 
a radiation dominated Universe, the
improved energy-momentum tensor for FRW space-time with torsion is
\beq\label{improved-energy-momentum}
{\cal T}_{ab}=\left(\begin{array}{cc}
\rho_0 & 0 \\ 0 & \frac {\rho_0} {3}\delta_{ij}
\end{array}\right) 
- \frac{b^2\pi G}{24}  T^6 \delta_{ab}
\eeq
with $\rho_0=\frac {\pi^2} {30} N_{eff} T^4$.
In fact both terms in (\ref{improved-energy-momentum}) are separately
conserved with the torsion-free connection.

Finally we observe that
the geometrical significance of non-zero $\t$ follows from 
the anti-symmetrised action of two co-variant derivatives on an arbitrary
vector field with components $U^a$,
\beq
[\nabla_a,\nabla_b]U^c =-\t^d{}_{ab}\nabla_d U^c +{R^c}_{dab}U^d.
\eeq
In addition to the algebraic (rotation) term involving the Riemmann
tensor there is a derivative term involving the torsion --- a
deficit displacement implying that parallelograms
generated by parallel transport do not close. 
The deficit displacement in Robertson-Walker space-time 
described here is
compatible with 3-dimensional rotational symmetry --- a vector
field with Robertson-Walker symmetries must have $U^i=0$ in which case
\bea
[\nabla_i,\nabla_j]U^0 &=&
-\t\,\epsilon_{ij}{}^k\nabla_k U^0 
\eea
and this vanishes if $U^0$ independent of position.
For any field compatible with the Robertson-Walker symmetries
space-like parallelograms close with the torsion
studied here. However they need not close for fields that do
not share the symmetries of the background metric.

The author wishes to thank the Perimeter Institute, Waterloo, Canada,
and the Dublin Institute for Advanced Studies for financial support.

\end{document}